\documentclass{ws-ijmpa}
\usepackage{amsmath}
\def\p{\partial}

\def\G{\Gamma}
\def\g{\gamma}

\def\de{\delta}
\def\De{\Delta}
\def\ld{\lambda}

\def\e{\eta}

\def\th{\theta}

\def\om{\omega}
\def\ov{\overline}
\def\s{\sigma}

\def\a{\alpha}
\def\ast{\alpha^*}

\def\pdellx'{\frac{\partial}{\partial x'}}
\def\pdellw'{\frac{\partial}{\partial w'}}

\newcommand{\be}{\begin{equation}}
\newcommand{\ee}{\end{equation}}
\def\bed{\begin{displaymath}}
\def\eed{\end{displaymath}}
\def\bea{\begin{eqnarray}}
\def\eea{\end{eqncrray}}
\def\[{$$}
\def\]{$$}
\newcommand{\beas}{\begin{eqnarray*}}
\newcommand{\eeas}{\end{eqnarray*}}

\begin{document}
\title{Confining QCD Model with Linear and 1/r potentials and Non-Propagating `Confion' Based on General Yang-Mills Symmetry}
\author{Jong-Ping Hsu \\
 Department of Physics, University of Massachusetts Dartmouth,\\ North Dartmouth, MA 02747-2300, USA\\
}

\date{October 2019}
\maketitle
{\small  A confining quantum chromodynamics (QCD) model is formulated on the basis of a new general Yang-Mills $SU_3$ symmetry.  The general Yang-Mills transformations involve arbitrary vector gauge functions $\om_\mu(x)$ and Hamilton's characteristic phase functions.   We derive fourth-order equations for new `phase fields', which predicts dual linear and Coulomb-like potentials for quark confinement. 
   The quantum of the `phase field' is called `confion'. The confion coupling strength turns out to be small, $g_s^2/(4\pi) \approx 0.2$, based on Cornell's results for charmonium.  Confions have indefinite energies.  However, such unphysical energies of confions cannot be detected because they are permanently confined within the  quark systems.  Furthermore, Green functions associated with confions do not propagate in space-time. }

\bigskip


\bigskip

\section{Introduction }
In previous works, we discussed a confining quark model with a general Yang-Mills symmetry and the groups $SU_3\times U_1$\cite{1,2,3}.  Their classical implications of static linear potential related to quark confinement and the late-time accelerated expansion of the baryon matter universe were discussed.

In this paper, we concentrate on the confinement properties of `gauge bosons' in the color $SU_3$ sector.  In such a confining QCD model, the key idea is the new generalized Yang-Mills (gYM) symmetry with arbitrary vector gauge functions and Hamilton's characteristic phase functions.\cite{4,5}  Such a new gYM $SU_3$  symmetry leads to fourth-order partial differential equation for the new `phase field' $H^a_\mu(x)$, which predicts dual static linear and Coulomb-like potentials for quarks.  The linear potential is generated by the quark current source $g_s \ov{q} \g_\nu(\ld^a/2) q$, while the Coulomb-like potential is generated by the phase field current source $\propto g_s \p^2[f^{abc} \p^\mu(H^b_\mu H^c_\nu)]$.  It is gratifying that the confion coupling strength turns out to be smaller than 1, $g_s^2/(4\pi) \approx 0.2 <1$, based on experimental results for charmonium.  This is interesting because the confining QCD model will be suitable for low energy physics and for perturbation theory.

 The quantum of the new phase field is called `confion'.    The Green function of the fourth-order confion equation suggests a very strange picture.  Namely, once there exists a source $\de^4(x)$, a constant field instantly appears everywhere inside the forward light cone, and there is a finite discontinuity on the light cone. In comparison, the invariant Green function in electrodynamics gives a delta function on the light cone and vanishes everywhere inside and outside the light cone. This property is related to the retarded potential in electrodynamics.\cite{6} 

Historically, the fourth-order field equation is usually considered unphysical because the dynamical system involves non-definite energy, otherwise there is no essential difficulty, according to A. Pais and G. E. Uhlenbeck, T. S. Chang and others.\cite{7,8}  However, in the present confining QCD model with general Yang-Mills symmetry, the massless confions are permanently confined in the quark system and, hence, their negative energies cannot be detected.   The confining QCD model suggests that there is no observable free confion.  Moreover, these `confions' should be treated as `off-mass-shell particles' in the external states and intermediate states of the S matrix.  Hence, they do not contribute to the imaginary part of physical amplitudes to violate unitarity.\cite{9,10}  
 
 We note that previous investigations of higher-order Lagrangians did not include gauge fields with dynamical $SU_N$ symmetry groups.  Indeed, the situation changes when the ideas of general Yang-Mills (gYM) symmetry and confinement are considered simultaneously.  It appears that for the gYM symmetry with simple 
 $(U_1)_{baryon}$ group, one could have a field theory with a finite fermion self-energy.  When this is applied to neutrinos, the result could have interesting implications to neutrino oscillations and dark matter because neutrinos must have non-vanishing masses\cite{11}. 

\section{Generalized Yang-Mills symmetry with arbitrary vector gauge functions}
Empirically, both linear and Coulomb-like potentials are necessary in order to understand the spectrum of charmonium, etc.\cite{12}
It is very unlikely that one can give a simple derivation of a linear potential from the usual second order gauge field equation with local source within the framework of relativistic local field theory.  However, if we generalize the usual gauge symmetry involving an arbitrary scalar gauge function  to a general Yang-Mills  (gYM) symmetry  with an arbitrary vector gauge function, one can derive dual  linear and Coulomb-like potentials from the fourth-order field equation, with the help of the Fourier transforms of generalized functions.\cite{13} 

For readers not familiar with general Yang-Mills symmetry, we briefly explain basic idea and equations.  General Yang-Mills (gYM) symmetry involves an arbitrary infinitesimal vector function $\om_\mu(x)$ and a new Hamilton's characteristic phase function $P(x)$ is defined as follows for quarks with color $SU_3$:
\be
q'(x) = e^{-iP}q(x)\approx (1- iP)q(x),    
\ee
$$
 \ov{q}'(x) \approx \ov{q}(x) (1+iP),   
 $$
\be
  P=L^b\left( g_s \int_{x'_o}^x dx'^\mu \om_\mu^b (x') \right)_{Le} \equiv g_s L^b P^b(\om,x), 
\ee
$$
  [L^a, L^b]= if^{abc} L^c, \ \ \  c=\hbar=1,  \ \ \ \   \e_{\mu\nu}=(+,-,-,-),
$$
where $P=P(\om,x)$, and $\om\equiv \om^b_\mu(x)$ are infinitesimal arbitrary vector functions (for simplicity) and $\ld^b=2L^b$ are Gell-Mann matrices.\cite{14}    Hamilton's characteristic phase function P is an action integral, which involves a fixed initial point and a variable end point.\cite{4,5}  
The initial point $x_0'$ in (2) is defined to be any fixed point, so that there is no contribution from it. It has no physical meaning.  If one chooses the variable end point x to be far away from the starting point, then of course the approximation (1) is not justified.    In this case, one uses finite transformations, which can be obtained from infinitesimal transformations in the usual way for Lie groups with, say, closed Lie algebras.  We have calculated consecutive general Yang-Mills transformations for $SU_N$ group.\cite{1}  As usual, these calculations are performed with infinitesimal transformations for simplicity.\footnote{   The calculations will be very complicated and highly non-trivial, if one uses finite transformations with elements of group $exp(-ig_s P^a L^a)$ in (1) and with the help of Baker-Campbell-Hausdorff formula.\cite{14}  The author would like to thank D. Fine for useful discussions. }   
 We stress that this new phase function P involves an infinitesimal `vector' gauge function $\om_\mu^a (x)$ and, hence, differs from the usual phase function (which involves a scalar gauge function) in the conventional gauge theories. This is crucial for general Yang-Mills symmetry.  For this new phase function to have unambiguous partial derivatives, we must impose a Lagrange equation (Le) to specify the path, just like Hamilton's characteristic function,\cite{4,5} which is a local function and, hence, compatible with local gauge theory. 

 The gYM transformations for the color $SU_{3}$ confining `phase fields' $H^a_\mu(x)$ are given by
\be
H'_\mu(x) =H_\mu(x) + \om_\mu(x)  -i[P(x), H_{\mu}(x)],    
\ee
 $$
 H_\mu =H_{\mu}^a L^a, \ \ \ \   
 $$
To see the equation  $Le$ in (2), let us consider the variation of $P$.\cite{4,5}  We have
$$
 \de P =g_s  \frac{\p \ov{L}}{\p\dot{x}^\ld}\de x^{\ld} 
$$
\be
 + g_s \left( \int_{\tau_o}^{\tau}\left(-\frac{d}{d\tau}\frac{\p \ov{L}}{\p \dot{x}^\ld}  
 + \frac{\p \ov{L}}{\p x^\ld}\right)\de x^\ld d\tau\right)_{Le},
\ee
where we write (2) in the usual form of a Lagrangian with the help of a parameter $\tau$,
\be
P=\left( g_s \int_{\tau_o}^\tau \ov{L}  \ d\tau\right)_{Le}, \ \ \    \ov{L}=\dot{x}^\mu \om_\mu^a (x) L^a, \ \ \  \dot{x}^\mu=\frac{dx^\mu}{d\tau}.
\ee
We require the paths in (4) to be those that satisfy the Lagrange equation $Le$, i.e., 
\be
-\frac{d}{d\tau}\frac{\p \ov{L}}{\p \dot{x}^\ld} + \frac{\p \ov{L}}{\p x^\ld}=0,    \ \ \  \ov{L}=\dot{x}^\mu \om_\mu^a (x) L^a.
\ee
 The integral in (4) vanishes and the Hamilton characteristic phase function $P$ in (2) is a well-defined function of x.  This property leads to an unambiguous relation\cite{4,5}
\be
 \p_{\mu} P^a=\om^a_\mu,  
\ee
which is a necessary relation for the general Yang-Mills symmetry. 

\section{`Phase field', its linear potential and Fourier transforms of generalized functions}

 As usual, the color $SU_{3} $ gauge covariant derivatives are  defined as
\be
 \De_{\mu} = \p_{\mu} + ig_{s}H_{\mu}^{ a} L^a.  
\ee
The $SU_{3}$ gauge curvatures $H^a_{\mu\nu}$ are given by
\be
[\De_\mu, \De_\nu]= ig_s H_{\mu\nu},  \ \ \ \   H_{\mu\nu}=H_{\mu\nu}^{ a} L^a,
\ee
\be
H^a_{\mu\nu}=\p_\mu H^a_\nu - \p_\nu H^a_\mu - g_s f^{abc}H^b_\mu H^c_\nu.
\ee

It follows from equations (1)-(10) that we have the following gYM transformations for $\p^\mu H_{\mu\nu}(x)$, and $\ov{q}\De_{\mu} q$:
\be
\p^\mu H'_{\mu\nu}(x)= \p^\mu H_{\mu\nu}(x)- i[P(x), \p^\mu H_{\mu\nu}(x)]
\ee
\be
 \ov{q}' \g^\mu \De'_{\mu} q'  =  \ov{q}\g^\mu  \De_{\mu} q  , 
 \ee
provided the restrictions  
\be
\p^\mu \{\p_\mu \om_\nu (x) - \p_\nu \om_\mu (x)\} - ig_s [\om^\mu (x), H_{\mu\nu}(x)] = 0
\ee
are imposed for (11) to hold.  Nevertheless, we still have infinitely many vector gauge functions $\om^a_\mu (x)$.   This constraint (13) is similar to that for gauge functions of Lie groups in the usual non-Abelian gauge theories.\cite{15}
  As usual, the gYM transformations have group properties.  We have calculated consecutive general Yang-Mills transformations for $SU_N$ group in ref. 1.  One can also verify that the general Yang-Mills transformations for, say, $SU_N$ reduces to the usual $SU_N$ transformations
 in the special case when the vector gauge function takes the form $\om^b_\mu(x)=\p_\mu \om^b(x)$.  To wit, we have $ P=L^b\left( g_s \int_{x'_o}^x dx'^\mu \p'_\mu \om^b (x') \right)_{Le} \equiv g_s L^b \om^b(x) $ in the special case.
  
Let us concentrate on the $SU_3$ sector of the gYM invariant Lagrangian, which is assumed to be
$$
L_{gYM} = \frac{L_s^2}{2}\left( \p^\mu H^a_{\mu\ld} \p_\nu H^{a\nu\ld}\right)
$$
\be
+i \ov{q}(x)\g^\mu\De_{\mu} q (x)- m_q \ov{q}(x)q(x), 
\ee
where the summation in those terms involving quarks, say, $m_q \ov{q}(x)q(x)$ is understood, i.e., $q(x)$ denotes the Dirac spinor for a quark field with mass $m_q$, and the sum over all 6 flavors and 3 colors, i.e., $\sum_{c=1}^{3} \sum_{f=1}^6 m_f \ov{q}^{cf} q^{cf} $.
Suppose we impose a gauge condition $\p_\mu H_a^\mu=0$,  the fourth-order gYM field equation takes the form
\be
L_s^2 \p^2 \p^2 H^a_{\nu} = J^a_\nu,  \ \ \ \    \p^\mu H^a_\mu=0,
\ee
\be
J^a_\nu= - g_s\ov{q} \g_\nu (\ld^a/2) q + g_s L_s^2 \p^2 [\p^\mu (f^{abc}H^b_\mu H^c_\nu)] +..... ,
\ee
where we have ignore some source terms which do not have the form $\p^2 [F(x)]$ and, hence, are not essential for discussions here.
Let us consider the static solution of  the time-component $H_0 (r)$ in (15) corresponding to the first source term $g_s\ov{q} \g_\nu (\ld^a/2) q $, which has the similar form of the charged current in electrodynamics.  Thus, as usual, we replace it by a static delta function $-g_s \de^3({\bf r})$, 
\be
 \triangledown^2 \triangledown^2 H_1 = - f_1 \de^3 ({\bf r}), \ \ \ \ \  
\ee
$$h
 H_1=H^a_0(r) , \ \ \ \ \    f_1= \frac{g_s}{L^2_s}, \ \ \ \  r=|{\bf r}|.
$$
To bring out the special mathematical properties in the solution to (17), we use the method of te Fourier transforms involving the generalized functions.\cite{13}  We define
\be
H_1(r) = \int e^{i{\bf k\cdot r}} \ov{H}_1(k) \frac{d^3 k}{(2\pi)^3}, \ \ \ \ \   k=|{\bf k}|,
\ee
\be
\ov{H}_1(k) = \int e^{-i{\bf k\cdot r}} {H}_1(r) d^3 r,
\ee
From (17) and (18), we have
$$
 \triangledown^2  \triangledown^2 H_1(r) = \int ( k^2)^2 e^{i{\bf k\cdot r}} \ov{H_1}(k) \frac{d^3 k}{(2\pi)^3}
 $$
 \be
 =- f_1\int e^{i{\bf k\cdot r}}\frac{d^3 k}{(2\pi)^3},
\ee
which gives
\be
\ov{H}_1(k) =\frac{- f_1}{(k^2)^2}, \ \ \ \ \ \    f_1= \frac{g_s}{L^2_s}.
\ee
From (18) and (21), we obtain the linear potential $H_1(r)$,
\be
H_1(r) = \int e^{i{\bf k\cdot r}} \frac{f_1}{(k^2)^2} \frac{d^3 k}{(2\pi)^3}=\frac{g_s r}{8\pi L^2_s}, 
\ee
where we have used the Fourier transform of generalized function,\cite{13}
\be
\int k^\ld e^{i{\bf k\cdot r}} d^3 k = 2^{\ld +3}\pi^{3/2} \frac{\G(-[\ld + 3]/2)}{\G(-\ld/2)} r^{-\ld-3}
\ee
where $\G(-1/2)=-2\sqrt{\pi}.$

Now let us consider the potential $H_2(r)$ generated by the second source term in (16).  Since it involves $\p^2$ acting on the boson source $g_s f^{abc}\p^\mu(H^b_\mu H^c_\nu)$ in the usual  $SU_3$ gauge theory, it is natural to interpret its static limit to be proportional to $\triangledown^2 \de^3({\bf r})$.  Thus, we may make the replacement 
\be
L_s^2 \triangledown^2 \triangledown^2 H_2 =  g_s L_s^2 \triangledown^2 \de^3({\bf r}), 
\ee
in (15), where we have used the only universal length scale $L_s$ to balance the dimensions on both sides of (24).  Thus, the potential $H_2(r)$ generated by the self-interaction of the phase field  satisfy
\be
 \triangledown^2 H_2(r) =  g_s \de^3({\bf r}), 
\ee
which has the same form as that in electrodynamics for Coulomb potential.  Thus, we also obtain a Coulomb-like potential in confining QCD with gYM symmetry,
\be
H_2(r) = - \frac{g_s}{4\pi r}.
\ee
Since the confions are permanently confined in the quark system, it is natural to consider these two static sources are located at the same position, so that $r$ in (26) is the same as that in (22).

  In confining QCD model with the gYM symmetry, the new phase field equation (15) implies two attractive potentials (22) and (26), which are very much different from that of the usual gauge fields.  The two sources of the phase field in (16) based on general $SU_3$ symmetry predict the following dual static potential energies $g_s H_0$,
 \be
g_s H_0 =g_s[H_1(r) + H_2(r)]=  \frac{g^2_s}{4\pi} \left[\frac{ r}{2 L_s^2} - \frac{1}{r}\right], 
\ee
$$
  \frac{g^2_s}{4\pi} \approx 0.2, \ \ \ \ \ \   L_s \approx 0.14 \ fm,
$$
where we have used Cornell's effective potential energies for charmonium.\cite{12}\footnote{The Cornell effective potential energy is $V(r)=(-\a_c/r)[1-(r/a)^2],$ where  $ \a_c\approx 0.2, \ a\approx 0.2 fm$.}

The dual potential energies (27) in the confining QCD model provide the necessary and clear mechanism for quark confinement and also for understanding the non-trivial charmonium spectrum, as shown by the comprehensive analysis of Cornell group
 (i.e., Eichten, Gottfried, Kinoshita,  Kogut, Lane, and Yan\cite{12}).   The Cornell result  provides a crucial experimental basis for the confining QCD model with the general Yang-Mills  $SU_3$ symmetry.  

Furthermore, it is gratifying that these results uniquely determine the values  of  the confion coupling strength $g_s^2/4\pi$ and the basic length $L_s$ in (27).
 The length $L_s$ is postulated to be a universal length and could play a role in particle-cosmology through the symmetry of baryon-lepton charges and the quark-lepton symmetry\cite{16}  
for theories based on general Yang-Mills symmetry.\cite{1} 
  The cosmic implications of the phase fields (with $(SU_3)_{color} \times (U_1)_{baryon}$) with the universal length $L_s$ for the late-time cosmic acceleration was discussed in previous works.\cite{1,2,3}

 \section{Confion Green functions --- Unusual and unobservable property of phase fields}
 
 The unusual space-time properties of phase fields or confions can be seen more clearly by comparing them with the electromagnetic fields or the photons.
 To see the physical propagation properties of the photons, one can consider the time-independent Green function in vacuum associated to inhomogeneous Helmholtz wave equation\cite{6}, i.e., $(\triangledown^2 + k^2) G_k(R)=-4\pi \de^3({\bf R})$.  One has, say, a divergent spheric wave from the source at the origin, $G_k^{(+)}(R)=e^{ikR}/R$.  For the time-dependent Green function, one has the well-known results,\cite{6}
 \be
 G^{(\pm)}(R, \tau) = \frac{1}{R} \de\left(\tau \mp R \right),
 \ee
 $$
 R=|{\bf R}|=|{\bf r} - {\bf r}'|, \ \ \  \tau=t-t'.   
 $$
 The argument in (28) implies an effect observed at the position ${\bf r}$ at time $t$ is caused by the action of a source at the position ${\bf r}'$ at an earlier time $t'=t-R, \ c=1$.  In other words, $t'-t$ is just the time of propagation of the interaction or disturbance from ${\bf r}'$ to the observation point ${\bf r}$\cite{6}.  
 
 For the space-time propagation property of the phase field $H^a_\mu(x)$ or the confions, we investigate the invariant Green function related to the phase field equation (15).  Its invariant Green function in vacuum takes the form
 \be
 \p^2_z \p^2_z D(z) = \de^4 (z), \ \ \ \ z^\mu = x^\mu - x'^\mu, \ \ \ \  \p^2_z=\frac{\p}{\p z^\mu} \frac{\p}{\p z_\mu},
 \ee
 for, say, the quark source.  The solution of (15) with the quark source can be expressed as
 \be
 H^a_\nu(x)=\int D(x-x') J^a_\mu(x') d^4 x'.
 \ee
 We define the Fourier transform $\ov{D}(k)\equiv \ov{D}(k_\s)$ of the Green function $D(z)\equiv D(z^\ld)$ by
 \be
 D(z)=\frac{1}{(2\pi)^4}\int \ov{D}(k) e^{-ik_\mu z^\mu} d^4 k.
 \ee
 Since $\de^4(z) =(2\pi)^{-4}\int d^4 k \ exp(-ik_\mu z^\mu)$, we have
 \be
 \p^2_z \p^2_z D(z) =\frac{1}{(2\pi)^4}\int \ov{D}(k) k^2_\ld k^2_\s e^{-ik_\mu z^\mu} d^4 k
\ee
$$
=\frac{1}{(2\pi)^4}\int d^4 k e^{-ik_\mu z^\mu}.
$$
Thus, we have
 \be
 \ov{D}(k)=\frac{1}{k^2_\ld k^2_\s}, \ \ \   k^2_\ld=k_\ld k^\ld = k_0^2 - \kappa^2, \ \ \ \kappa=|{\bf k}|,
 \ee
where $ \ov{D}(k)$ is proportional to the `confion propagator' in the  rules for Feynman diagrams in the confining QCD. Using (31) and (33), we solve the Green function $D(z)$ by carrying out the contour integration on the complex $k_0$ plane,\cite{6}
 \be
 D(z)=\frac{1}{(2\pi)^4}\int e^{-i{\bf k\cdot z}} d^3 k\int_{-\infty}^{\infty} dk_0 \frac{e^{-ik_0 z_0}}{(k_o^2 - \kappa^2)^2}.
 \ee
 The integrand has two poles of order 2 at $k_0=\pm \kappa$.  Invariant Green functions $D(z)$ are obtained by choosing different contours of integration relative to the two poles of order 2.  Similar to the retarded Green function with $z_0 >0$ in electrodynamics,\cite{6} we use Cauchy theorem and obtain
 \be
 \int_{-\infty}^{\infty} dk_0 \frac{e^{-ik_0 z_0}}{(k_o^2 - \kappa^2)^2} =\frac{-\pi}{\kappa^3}[sin(\kappa z_0) - z\kappa cos(\kappa z_0)],
 \ee
for $z_0 >0$.  The invariant Green function $D(z)$ can be expressed as
$$
D(z^\ld)=\frac{1}{(2\pi)^4}\int e^{-i{\bf k\cdot z}} d^3 k \frac{\pi}{\kappa^3}[ z\kappa cos(\kappa z_0) - sin(\kappa z_0)],
$$
\be
=\int_o^\infty {d\kappa}\left[\frac{z_0 sin(\kappa z) cos(\kappa z_0)}{(2\pi)^2\kappa z}-\frac{sin(\kappa z) sin(\kappa z_0)}{(2\pi)^2\kappa^2 z}\right],
\ee 
 where $z=|{\bf z}|$.  Finally, we obtain the following `retarded' confion Green function $D(z^\ld)$
 \be
 D(z^\ld)=\frac{1}{16 \pi} - \frac{1}{8\pi}=-\frac{1}{16 \pi}, \ \ \  z_0=z > 0,
 \ee
 \be
 D(z^\ld)=0 - \frac{1}{8\pi}=-\frac{1}{8 \pi}, \ \ \  z_0>z > 0,
 \ee
 \be
 D(z^\ld)=\frac{z_0}{8 \pi z} - \frac{z_0}{8\pi z}= 0, \ \ \  z>z_0 > 0.
 \ee
 
 For comparison with (37)-(39), the retarded Green function $D_r(z^\ld)$ in electrodynamics is\cite{6}
 \be
 D_r(z^\ld)=\frac{\theta(z_0)}{4\pi R} \de(z_0 - z)
  \ee
 $$
=\frac{1}{8\pi R} \de(x_0 - x'_0 -R). \ \ \ z_0=z>0, 
 $$
 which vanishes for $z_0>z>0$ and $z>z_0>0$.  The Fourier transform of $D_r(z^\ld)$ in (40) with respect to $z_0$ is
 \be
\int_{-\infty}^{\infty} e^{ik_0 z_0} D_r(z^\ld) dz_0= \frac{1}{4\pi R} e^{i\kappa R}
 \ee
 $$
 =\frac{1}{2\pi} \th(z_0)\de(z_\ld z^\ld), \ \ \ \ \ \ \ \  k_0=\kappa,
 $$
 which is the (photon) propagating Green function of outgoing waves with the speed of light.\cite{6}
 
 For comparison, the Fourier transform of the confion Green function (38) with $z=0$ and $z_0 > 0$ is as follows:
 \be
 \int_{-\infty}^{\infty} e^{ik_0 z_0} D(z^\ld) dz_0=  \int_{0}^{\infty} e^{ik_0 z_0} D(z^\ld) dz_0
  \ee
 \be
 =\frac{-1}{8}\left[ \frac{i}{\pi k_0} +  \de(k_0)\right], \ \ \ \ \ \ \    z_0 > 0,
 \ee
where we have used the Fourier transform of generalized function\cite{13}
\be
\int_{-\infty}^{\infty} \th(z_0) e^{iz_0 k_0} dz_0= \frac{i}{k_0} + \pi \de(k_0).
\ee 
The result (43) shows that the Green function associated with phase field $H^a_\mu(x)$ in the confining QCD model contains a constant $i/k_0$ and a singularity $\de(k_0)$, which does not have simple physical interpretation of propagation.  

These results (37)-(39) and (43) indicate that the phase field $H^a_\mu(x)$ does not propagate in space-time because it does not dependent on space and time, in sharp contrast to the out going waves with the amplitude $\propto 1/R$ as shown in (41) in electrodynamics.  The invariant Green functions (37)-(39) with $z=0$ suggests a simple new and strange picture:  Namely, once there is the source $\de^4(z)$ in (29), a constant phase field instantly appears everywhere inside and on the forward light cone.  Such a property appears to be interlocked with the confining potentials (27) in the confining QCD model with the gYM symmetry.  These properties are consistent with our interpretation that a confion does not propagate in vacuum as a free particle, in sharp contrast to the photon.

\section{Discussions}

The pure phase field $H^a_\mu$ by itself cannot have confinement property because the structure of the second source term of phase fields in (16) leads to Coulomb-like potential rather than the linear potential.  It is the quark source, i.e., the first term in (16), that enables the phase field to produce the linear potential in the confining QCD model.

 It appears that dual liner and Coulomb-like potentials are characteristic of general Yang-Mills $SU_3$ symmetry, as we have demonstrated in equations (15)-(26) for the static case.  In particular, equation (22) involves the Fourier transform of a generalized function $1/(k^2)^2$ and, hence, suggests that we may picture the linear potential as the exchange of one virtual confion between two quarks, where the confion satisfies a fourth-order differential equation. 
 
In summary, the most interesting results of the confining QCD model with the new general Yang-Mills symmetry are as follows:

(a) It explicitly provides a confining mechanism, i.e., the dual linear and Coulomb-like potentials (27) for quarks. 

(b) The confion is not detectable because it cannot propagate as a free particle, as shown by the solutions (37)-(39) of  the confion equation (15).  

 Furthermore, these results of the confining QCD model are supported by experiments and comprehensive analysis and results of the Cornell group\cite{12}.

 The work was partially supported by the Jing Shin Research Fund of the UMass Dartmouth Foundation. 

\end{document}